\RequirePackage{fix-cm}
\documentclass[smallextended]{svjour3}       % onecolumn (second format)
\smartqed  % flush right qed marks, e.g. at end of proof
\usepackage{graphicx}
\usepackage{eurosym}
\usepackage{siunitx}
\usepackage{floatrow}
\usepackage[table]{xcolor}
\begin{document}

%\title{Individualized Flexible Mitral Valve Models With Chordae Tendineae Enhance Education, Training and Planning For Reconstructive Surgery

\title{Flexible and Comprehensive Patient-Specific \\ 
Mitral Valve Silicone Models with Chordae Tendinae Made From 3D-Printable Molds
%Enable Surgical Training
%Individualized Flexible Mitral Valve Models With Subvalvular Apparatus Foster Training of Surgical Techniques for Reconstructive Surgery
%\title{Comprehensive Surgical Training Simulator for Minimally-Invasive Mitral Valve Reconstruction 
%Patient-Specific Silicone Mitral Valve Models Enable Surgical Skills Training 
%Patient-Specific Silicone Mitral Valve Phantoms For Surgical Dry Lab Training
% BVM: Elastic Mitral Valve Silicone Replica Made from 3D-Printable Molds Offer Advanced Surgical Training
%AUSKOMMENTIERT
\thanks{The research was supported by the German Research Foundation DFG project 398787259, DE 2131/2-1 and EN 1197/2-1.}
%Grants or other notes
}

%BVM: Elastic Mitral Valve Silicone Replica Made from 3D-Printable Molds Offer Advanced Surgical Training

\titlerunning{Flexible Silicone Mitral Valve Models}        % if too long for running head

\author{Sandy Engelhardt, Simon Sauerzapf, Bernhard~Preim, Matthias~Karck, Ivo~Wolf, Raffaele~De~Simone
}

\authorrunning{Engelhardt et al.} % if too long for running head

\institute{Sandy Engelhardt \and Simon Sauerzapf \and Ivo Wolf \at
              Faculty of Computer Science, University of Applied Sciences Mannheim, Germany \\
              \email{s.engelhardt@hs-mannheim.de}           %  \\
           \and
           Matthias Karck \and Raffaele De Simone \at
           Department of Cardiac Surgery, University Hospital Heidelberg, Germany 
             \and
             Sandy Engelhardt \and Bernhard Preim \at
             Faculty of Computer Science, Magdeburg University, Germany
             %\& Research Campus STIMULATE, Magdeburg University, Germany
}

\date{Received: date / Accepted: date}

\maketitle

\begin{abstract} %200-300 words

\textit{Purpose}
Given the multitude of challenges surgeons face during mitral valve repair surgery, they should have a high confidence in handling of instruments and in the application of surgical techniques before they enter the operating room. Unfortunately, opportunities for surgical training of minimally-invasive repair are very limited, leading to a situation where most surgeons undergo a steep learning curve while operating the first patients.

\textit{Methods}
In order to provide a realistic tool for surgical training, a commercial simulator was augmented by flexible patient-specific mitral valve replica. 
In an elaborated production pipeline, finalized after many optimization cycles, models were segmented from 3D ultrasound and then 3D-printable molds were computed automatically and printed in rigid material, the lower part being water-soluble. After silicone injection, the silicone model was dissolved from the mold and anchored in the simulator. 

\textit{Results}
To our knowledge, our models are the first to comprise the full mitral valve apparatus, i.e. the annulus, leaflets, chordae tendineae and papillary muscles. Nine different valve molds were automatically created according to the proposed workflow (seven prolapsed valves and two valves with functional mitral insufficiency). From these mold geometries, 16 replica were manufactured. A material test revealed that \textit{Ecoflex\textsuperscript{TM} 00-30} is the most suitable material for leaflet-mimicking tissue out of seven mixtures. Production time was around 36h per valve.Twelve surgeons performed various surgical techniques, e.g. annuloplasty, neo-chordae implantation, triangular leaflet resection and assessed the realism of the valves very positively.  

\textit{Conclusion}
The standardized production process guarantees a high anatomical recapitulation of the silicone valves to the segmented models and the ultrasound data. Models are of unprecedented quality and maintain a high realism during haptic interaction with instruments and suture material.

%\and minimally invasive mitral surgery \and mold creation}
\end{abstract}

\section{Introduction}
\label{sec:intro}

%Idee fuer klinisches Paper: Siehe Kenngott et al Einleitung
Reconstructive mitral valve surgery is a demanding surgical sub-speciality due to variation in valve pathologies, required dexterity and tedious acquirement of surgical techniques. Mastering the technical skills to successfully performing needs years of training. 
At the same time, opportunities to practice are limited and most of the training is conducted by assisting and performing real surgeries, following the surgical principle of ``see one, do one, teach one'' \cite{Kotsis2013}. 

% review
%Despite the fact that mitral valve repair is superior to the alternative mitral valve replacement for degenerative pathologies \cite{DeBonis2016}, many surgeons are still deterred from adapting this procedure because of the mentioned challenges \cite{Sardari}. This is underlined by a study from Bolling et al. in 2010 \cite{Bolling2010}, who has reported that the median number of operations per surgeon in the US is considerably low with 5 cases per year. 
%\textcolor{red}{Not surprisingly, higher surgeon volume was independently associated with an increased probability of mitral valve reconstruction in their study \cite{Bolling2010}.} 

Especially minimally-invasive procedures are considered as very complicated. It comprises a 5-6~cm incision in the intercostal space, which is referred to as a right lateral thoracotomy. 
The space constraints and special long-shafted instruments require surgeons to operate differently than in open surgery, e.g.\ the needle must be fixed in other angles to the needle holder as usually, which makes the stitching process more difficult. Additionally, knot tying is performed with a knot pusher, which can easily result in an air knot if done inappropriately.
%Surgeons may use a video endoscope for visual guidance. 
According to Holzhey et al. \cite{Holzhey483}, who analyzed a total of 3895 operations by 17 surgeons performing their first minimally invasive surgery of the mitral valve, the typical number of operations to reach a high quality level was between 75 and 125. In addition, more than one of such an operation per week was necessary to maintain good results. They identified a clear need for good mentoring and training utilities in the learning phase.

%To date, training for surgery in general can be conducted on \textit{ex-vivo} organs (so-called `wet labs'), on physical simulators equipped with organ replica or on virtual simulators (so-called `dry labs'). Unfortunately, restricted tissue lifetime and organizational efforts limit the realization of wet lab courses in clinical practise.
%Virtual simulators overcome these drawbacks, but underlie restrictions in simulation of tissue properties, means for interaction and haptic feedback.
Besides `wetlabs' or virtual simulators, physical phantoms provide options for surgical training. However, only the latter has the unique ability to bring the patient's anatomy into physical reality as a model. 
Additionally, it provides excellent opportunity for dexterity training with authentic instruments and suture material \cite{Kenngott2015}, while offering materials that are pliable enough to be cut and sewn. %The benefit might be especially high if patient-specific replica are used, which are capable of reflecting individual pathology. 
Moreover, such models have the advantage that they could be used before the actual surgery takes place to rehearse the procedure and to discuss with colleagues without time constraints.
%before the actual operation takes place. 
%terms of accurately planning and training critical steps of the reconstructive method applied in real surgery.

Mitral valve repair is a highly experienced-based surgery and techniques are chosen according to the surgeon's preference. Various surgical repair techniques exist (e.g. annuloplasty, leaflet plasty, chordae plasty), which may address different parts of the valve \cite{Praet2018}. A patient-specific mitral valve phantom consisting of the annulus, leaflets and the subvalvular apparatus would allow the performance of the full spectrum of repair techniques and may be a valuable tool for comparison of current `competing' techniques like triangular leaflet resection and neo-chordae implantation. 
The measurement of chordae length as well as implantation and knot tying of neo-chordae at their determined length are procedures considered most crucial in mitral valve repair today. Performing this procedure with a high confidence after training on a phantom can avoid a sub-optimal surgical outcome such as re-occurrence of valve leakage or restricted leaflet motion.

In a previous work \cite{EngelhardtBVM2018}, we demonstrated the feasibility for producing elastic 3D-replica consisting of mitral annulus and the two mitral leaflets. Using our customized software for valve modelling \cite{EngelhardtFIMH} allowed us to extract models with high fidelity from 3D echocardiography, which is one elementary key to the work at hand. %Automated negative mold prototyping  to cast silicone mitral valves. 
The production pipeline consisted of several comprehensive steps, e.g. image acquisition, 3D-segmentation of the mitral valve, computation of negative molds and material casting.
In this work, we extended the models to incorporate the complete mitral valve apparatus, which includes the complex branching structure of the chordae tendineae and the papillary muscles.   %pseudo-chordae distribution
For this task, we had to completely revise the molding and casting concept and optimized it in several development cycles, as it is not straightforward to produce a negative mold that can be easily disassembled after silicone casting due to the complex nature of the valvular shape. %Furthermore, the current software had to be revised to be able to capture the papillary muscles and a concept for generation of a chordae distribution was developed. 
In this paper, we present the revised production process, novel generic computer-aided design parts, a material study and a larger user study conducted with trainee and expert surgeons. The goal of this work was to develop unique valve models of unprecedented quality that combine highly realistic anatomy and material properties suitable for surgical training in a cost-effective production process to motivate dissemination in the medical community. 
%Ausserdem neu: Improved Casting process (luer lock), novel CAD parts.

%Bildgebung 
%Verschiedenste Studien stellten aber bereits ein breitesSpektrum an Analysemöglichkeiten der 3D-TOE heraus, bspw. für die erweiterte Krankheitscharakterisierung[3,10], die chirurgische Planung [11] oder die Risikobeurteilung für Systolic Anterior Motion [12].

% Details zur OP
%All diese Faktoren führen dazu, dass heute noch in vielen Nicht-Expertenzentren deutlich zu viele Klappenersatz-Prozeduren auch bei degenerativen Erkrankungen durchgeführt werden [6], obwohl die Rekonstruktions-Remodellierungs-Chirurgie mit sehr guten postoperativen Ergebnissen und einer schnellen Rekonvaleszenz einhergeht [2].

\section{Related Work}
%https://www.fasebj.org/doi/abs/10.1096/fasebj.31.1_supplement.391.5

%https://www.laz3d.com/thoracic-surgery-models.html

%Second, to realistically simulate valve repair, “valve-like” materials must be pliable and elastic enough to be cut and sewn.

%To our knowledge, virtual simulators for training of mitral valve repair are not available. Therefore, in the following review we will focus on physical simulators for mitral valve repair. 
A handful of companies provide physical mitral valve simulators. The \textit{Mitral valve holder} %\cite{lifelikebio}
(LifeLike BioTissue Inc., Ontario, Canada) or the \textit{MA-TRAC High Fidelity Minimally invasive Mitral Valve Repair simulator} %\cite{maastricht} 
(Maastricht Trading Company, Inc., Maastricht, The Netherlands) are equipped with mitral valves made of silicone.
Others companies provide 2D valves models made of felt, such as the \textit{MICS MVR Simulator} (Fehling Instruments GmbH \& Co. KG, Karlstein, Germany). %\cite{fehling}.
However, such valves are generic, lack patient-specific properties and are therefore not suitable to address strategies for particular valvular conditions. 

Existing works already dealt with direct 3D-printing of the mitral valve using acrylonitrile butadiene styrene (ABS) plastic material \cite{2611-ref02}. Such stiff material showed benefits for surgery planning and surgical teaching, but it is certainly inadequate for dexterity training.
Vukicevic et al. \cite{2611-ref03} also used direct printing, but employed multi-material elastomeric TangoPlus materials (Stratasys, Eden Prairie, Minnesota, USA) that were compared to freshly harvested porcine leaflet tissue. All TangoPlus varieties were less stiff than the maximum tensile elastic modulus of porcine mitral valve tissue. 
%The work mainly addresses the field of  interventional cardiology and
Furthermore, the mesh creation required a lot of manual input and chordae tendineae were not included in the final rapid prototyping models. %Other work for cardiac surgery training also employed direct printing \cite{Yamada2017}, but the materials u.

%Regardless of the tissue stiffness, surgical eligibility of the material in terms of realistic stitching properties also needs to be evaluated. 
%porose

Scanlan et al. \cite{Ilina2017SPIE,Scanlan2017} proposed a method for creating 3D-printable molds, which can be filled with an elastic material such as silicone to create flexible and tear-resistant pediatric leaflet models. Beyond that, they compared direct printing with flexible material. Molding required more time and labor than directly printed models, but permits materials that better simulate real tissue and that are more economical at scale. Furthermore, direct printing of flexible material usually requires expensive 3D printers and materials. 
Surgeons reported good tissue properties for the cast valves (realistic flexibility, cuts and holds sutures well without tearing) from their personal experiences and considered the silicone models to be useful for surgical planning and training. However, their valve models did not include the subvalvular apparatus. 

In a very recent medical review, Ginty et al.\ \cite{Ginty2018} describe a general guidance for creation of dynamic valve models usable in a flow simulator. The authors provide an overview of the production pipeline (imaging, segmentation, 3D printing, material casting) and discuss viable options for each step. Moreover, they present an approach, for which two interesting points shall be highlighted: They just printed the lower impression of the valve and not a full mold. Silicone was painted on the printed surface in a thin layer, which then cured to the shape of the valve. Furthermore, they incorporated braided nylon
fishing line strings into the leaflets to mimic some strings of the chordae tendineae. With the mentioned steps, it seems to be difficult to obtain results of the same quality when multiple valves are produced from the same mold (e.g. two models could have different thicknesses or varying chordae attachments). The phantoms do not seem to include papillary muscles.
%Apart from that, the group around Sardari et al. \cite{Sardari} showed the feasibility of a completely manual approach for production of a single valve model.

% Zusammenfassender Abschnitt:
To sum up, none of the current approaches allow for  production of comprehensive and flexible models consisting of all anatomical substructures of the valve within the scope of a standardized production process that maintains the quality over different fabrications.

%the ideal saddle shape of the annulus was not reflected in the models.

%The shape of structures can be accurately modeled using direct 3D printing, but currently available soft printing materials may still be too rigid to imitate human tissue. 
 
%\begin{figure}
%  \centering
%    \includegraphics[width=0.2\textwidth]{FehlingValve.png}
%     \caption{Generic felt valve}
%     \label{fig:felt}
%\end{figure}

%Yamada et al. \cite{Yamada2017}
%Minimally Invasive Cardiac Surgery

%Plus andere arbeiten.

\section{Materials and Methods}

\begin{figure}
  \centering
    \includegraphics[width=\textwidth]{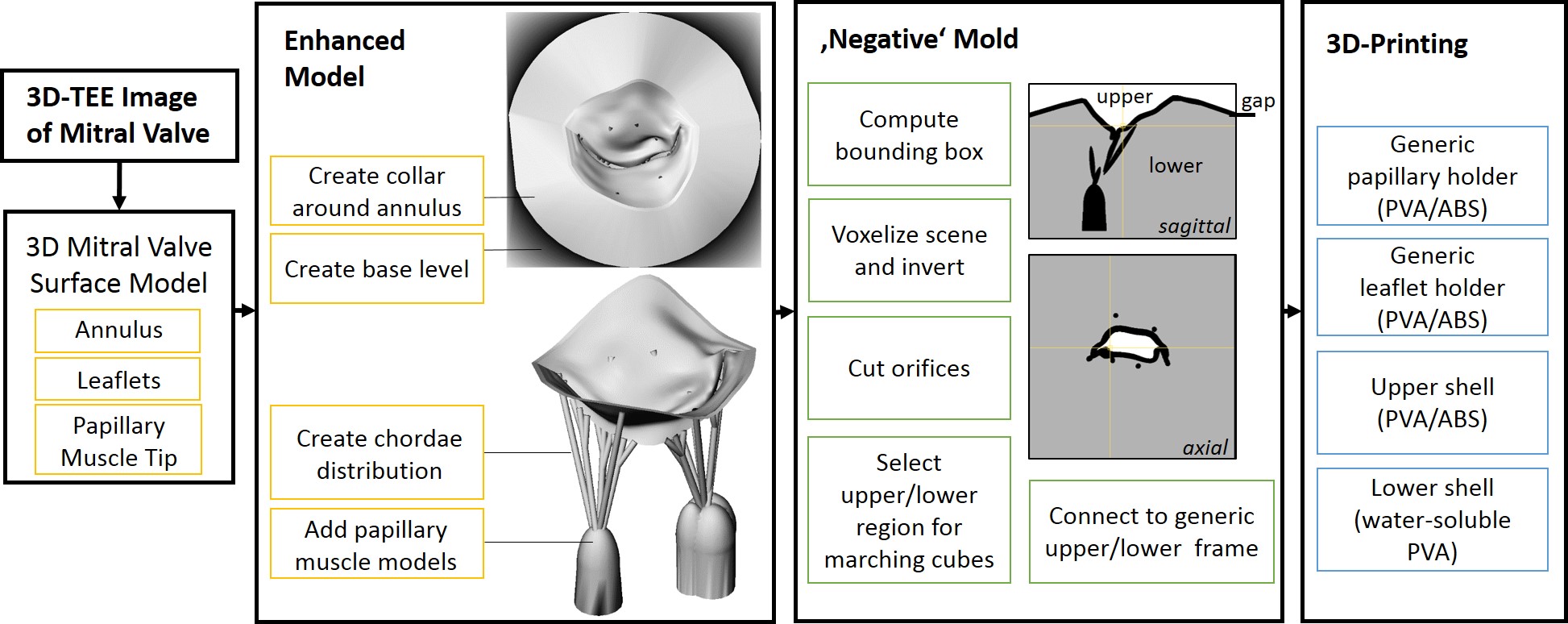}
    \caption{Main steps for creation of mold parts and holders.}
    \label{Fig:Patientspecific}
\end{figure}

Our developments are based on the commercially available \textit{MICS MVR Simulator}, %(Fehling Instruments GmbH \& Co. KG, Karlstein, Germany)
which resembles intraoperative space constraints. The flat valve dummies, that are delivered with the simulator, are concealed inside this simulator, anchored on a bar and can be reached via a small circular access of 50~mm diameter. 
%The view on the operation area can be direct, but is encouraged to be indirect over an integrated low budget video camera that streams to an external display. 
%The object support for annulus and mitral valve dummy and papillary muscle dummy can move freely in the guide rail, and thus the distance to each other and to the access is variable.
%A basic set of long shafted instruments is included in the supply of the simulator and comprises needle holder, forceps, valve scissors, knot pusher, nerve hook. 
%The simulator's textile valve consists of two simple sheets as a dummy for mitral annulus and leaflets. Furthermore, the papillary muscles are represented by two tube-shaped objects. Chordae tendineae are not present. 
Note that the geometry of the valve is neither patient-specific nor three-dimensional. 
%The surgeon can more or less perform annuloplasty by implanting a felt ring, a triangular resection at the posterior leaflet or a chordal replacement.
Given the generic nature of the dummy, the surgeon can perform some reconstruction techniques, but the ``created'' dummy geometry is meaningless from an anatomical and functional point of view, not fostering individualized therapy.

To offer advanced means for training and preparing of surgery, we extended the existing simulator by flexible patient-specific mitral valves generated from medical images.
We identified the following requirements (R2 extends our previous requirements \cite{EngelhardtBVM2018}; R1-R3 refers to position and shape; R4-R7 to material, costs and reproducibility):
%\begin{itemize}
%\item[a)]{Position and Shape:}
\begin{itemize}
\item{R1: The model must be fixable into the MICS MVR simulator to provide a
realistic port-to-valve relation.}

\item{R2: The model should consist of all parts of the valve: annulus, anterior and posterior leaflets, papillary muscles and chordae tendineae to allow for training of the full spectrum of repair techniques.}

\item{R3: The leaflets should reflect end-systolic shape of patient-specific valve
to incorporate the morphology of potential prolapsing leaflet segments at the moment of valve closure.}
%\end{itemize}

%\item[b)]{Material, Costs and Reproducibility:} 
%\begin{itemize}

\item{R4: Stitching properties of the material should be realistic such that 
needles can be inserted and sutures pass through with similar tissue resistance as in real surgery.}

\item{R5: Cutting properties of the materials should be realistic in terms of resistance.}

%\item{R6: The used material should be isoechogenic, in order to be able to investigate the effects of individualized therapy in a flow simulator.}

\item{R6: Common materials for low-cost 3D-printing should be used
to facilitate reproducibility of the approach by other groups.} %ok

\item{R7: Total production time should be less than 2 days and the process as automated
as possible to facilitate integration into the routine clinical use.}
%\end{itemize}
\end{itemize}
In the following, we elaborate on the automatic mold creation process. In short, from given 3D virtual models, a negative shape is computed that is connected to generic CAD-parts to build a sealed block for material insertion. %The mold consists of an upper and lower part and a leaflet holder, which is placed inside the mold. 

\subsection{Medical Imaging and Mitral Valve Modeling}
3D polygonal surface models of the mitral valve are obtained from transesophageal echocardiography (TEE) using customized software \cite{EngelhardtFIMH}. The software was extended such that the position of papillary muscle heads can be localized. The mitral valve's annulus, mitral leaflets and the position of the papillary muscle heads are segmented from the end systolic time frame (R3). This information is used to create an `enhanced model' (Fig. \ref{Fig:Patientspecific}):
The fine branching structures of the chordae tendineae are not completely visible on current clinical image modalities \cite{Ginty2018}, therefore, a chordal distribution is automatically created. 
It consists of primary and secondary chordae tendineae, taken the leaflet geometry and the papillary muscle tip as reference. % (Fig. \ref{fig:chordae}).
We employed anatomical descriptions as guide for creating the distribution. 
For the primary chordae, points at the free edge of the leaflet surface are connected to the respective papillary muscle tip. The number of main chordal strings is an adjustable parameter in the software and the insertion points are automatically adopted accordingly. The chordal strings are equidistantly distributed along the free edge of each leaflet and connected to the anterolateral or posteromedial papillary muscle. Each main chordal string gives rise to an additional left and right branch randomly attached at the mid of the string or higher. % (Fig. \ref{fig:chordae}). 

%\begin{figure}
%\floatbox[{\capbeside\thisfloatsetup{capbesideposition={right,top},capbesidewidth=5cm}}]{figure}[\FBwidth]
%{\caption{Generated chordae distribution consisting of primary (some marked in green, orange, yellow) and secondary strings (some marked in blue). The point of insertion at the papillary muscle tip is randomly spread around the head of the papillary muscle and the height of the branching is randomly determined.}\label{fig:chordae}}
%{\includegraphics[width=6cm]{Chordae1.jpg}}
%\end{figure}

%\begin{figure}
%  \centering
%    \includegraphics[width=0.5\textwidth]{Chordae1.jpg}
%     \caption{Generated chordae distribution consisting of primary (some marked in green, orange, yellow) and secondary strings (some marked in blue). The point of insertion at the papillary muscle tip is randomly spread around the head of the papillary muscle and the height of the branching is randomly determined.}
%     \label{fig:chordae}
%\end{figure}

\begin{figure}
  \centering
    \includegraphics[width=\textwidth]{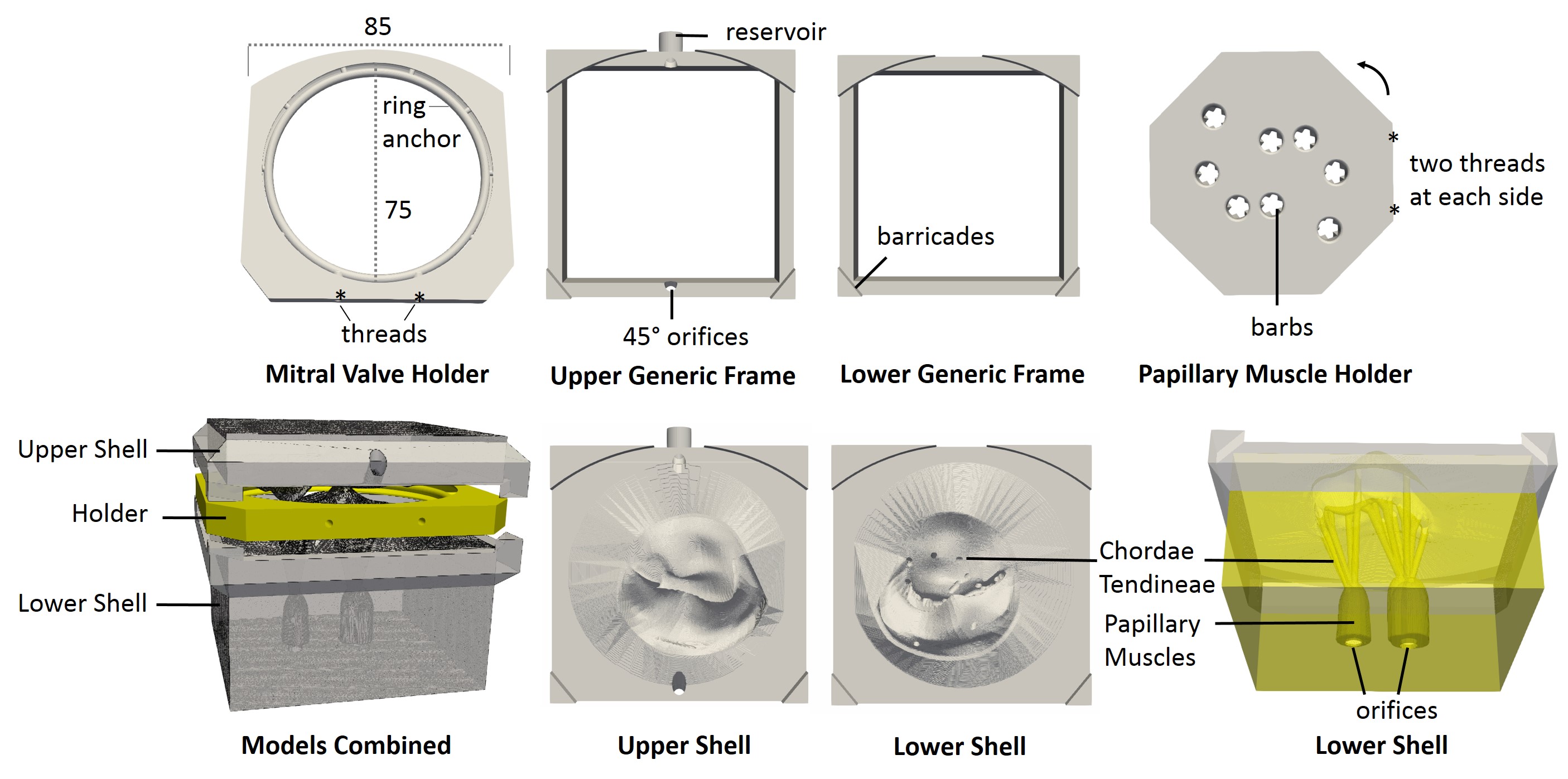}
     \caption{Patient-individual valve surfaces are incorporated into the generic frames.}
     \label{fig:CAD}
\end{figure}

\subsection{CAD-Modelling of Generic Parts}

 %software to adapt the segmented mitral valve geometry into an appropriate fixture for securing into the 
 %The simulator requires
%Describe model of papillary muscle here
Two papillary muscle models were created in the freely available software Blender v2.79 and were aligned with the individual papillary muscle tips. The anterolateral head (diameter of 8.7~mm) has a single head, whereas the posteromedial papillary muscle has three heads (diameter of 11~mm). They both extend towards the lower end  to anchor them in a specifically designed papillary muscle holder (Fig. \ref{fig:CAD}). 
Other parts are modeled once in the CAD software Autodesk Fusion 360 (v.2.0). The parts either serve as a common frame for the mold parts (referred to as \textit{upper} and \textit{lower shells} or \textit{mold parts}) or as a fixture for securing of the valve into the simulator (referred to as \textit{leaflet holder} and \textit{papillary holder} (R1)). The first row in Fig.\ \ref{fig:CAD} provides an overview of the generic parts. 
The leaflet holder incorporates a circular hole with a ring anchor, which will be enclosed by silicone when placed in between the two shells. 
This concept ensures that the holder remains permanently connected to the  valve model. %This holder is placed inside of the mold parts, which enclose it completely.
%The holder has two small threads canals for securing the part with screws in the simulator.
The leaflet holder and the shells have undergone several modifications in an iterative process during the scope of this project in comparison to our previous work \cite{EngelhardtBVM2018}: the frames have several 4~mm orifices that allow for injection of silicone by a Luer lock syringe (previously, the material has been cast and not injected); barricades were added to the frames to secure them when they are stuck together; the size of the holder with ring anchor was increased to accommodate bigger valves;  $90^\circ$ overhangs were removed to cut down support structures. 

\begin{figure}
\floatbox[{\capbeside\thisfloatsetup{capbesideposition={right,top},capbesidewidth=2cm}}]{figure}[\FBwidth]
{\caption{Injection molding.}\label{fig:Spritz}}
{\includegraphics[width=9cm]{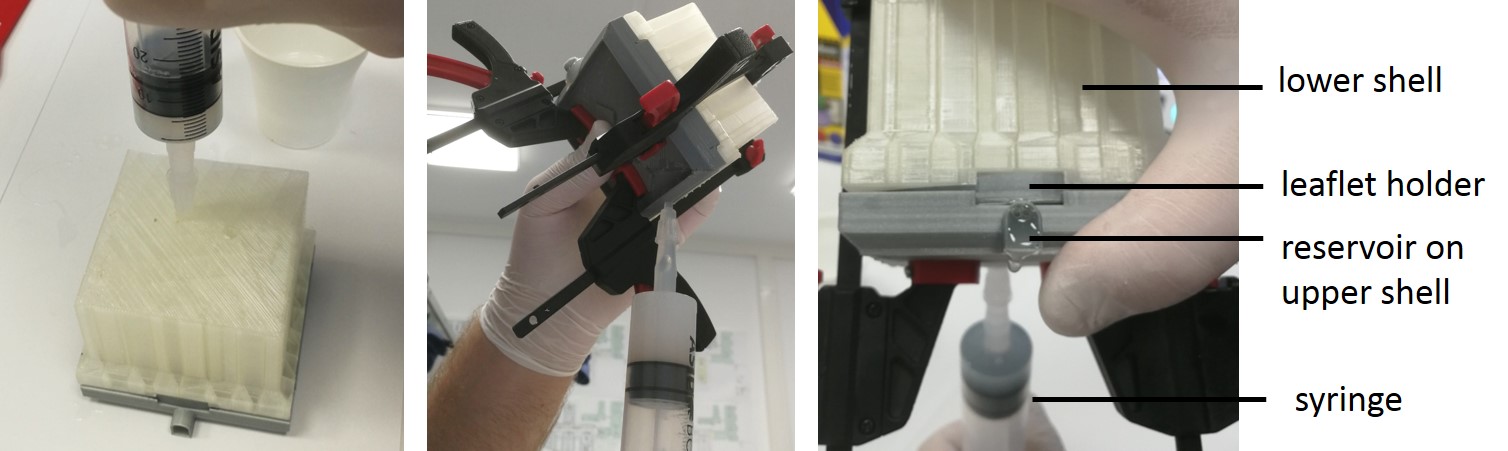}}
\end{figure}

%The papillary holder (cf. Fig. \ref{fig:CAD}) was designed once and is usable for all valve models that are produced. It has an octagonal structure and a random arrangement of holes with barbs to fix the papillary muscles. This holder can be mounted in an arbitrary orientation onto the bar of the simulator.
% EVTL WIEDER EINKOMMENTIEREN
%The upper shell has two $45^\circ$ orifices at two opposite sides. One orifice is equipped with a small reservoir that is completely filled with silicone. The position of the reservoir will be the highest point during solidification and small amounts of silicone might be sickening into the mold while last air bubbles ascend.
% EVTL WIEDER EINKOMMENTIEREN
%An endoscopic port attachment has been created for the simulator to insert an endoscopic camera that produces high quality images (Fig. XYZ).  
%Angle Element for simulator?

\subsection{Generation of Upper and Lower Casting Mold Part}

In a next step, the mesh of the generic frames are extended with the patient-specific valve surfaces (Fig. \ref{Fig:Patientspecific}). The software toolkit MeVisLab 3.0.2 and the C++ software library Visualization Toolkit (VTK) 8.0.1 were used for this task. 
First, the valve model's position and orientation is normalized. Then, the gaps between generic frame and valve model are closed by triangle strips. %on a base level formed on the level of the mean annulus plane. 
After obtaining this ``enhanced surface model'' (Fig. \ref{Fig:Patientspecific}), leaflets of a certain thickness had to be created from the provided surfaces. The previously used method VTK implicit function \cite{EngelhardtBVM2018} delivered a lot of artifacts due to complicated triangular topology and multiple surfaces. To circumvent this issue, the MeVisLab module \textit{VoxelizeInventorScene} has been employed, which enables the computation of a voxel representation of the scene, i.e.\ a 3D bounding box volume that masks out the valve. As voxel size, we chose (0.01~mm)$^3$. A distance in $\mathrm{mm}$ to the leaflet needs to be set, which adds a thickness to this otherwise flat surface representation. We set this parameter to 0.8, which extends the surface to both sides. 
The resulting image mask content was separated into a lower and an upper region representing the patient-specific part of the shells. Each part was connected to the respective generic frame yielding the final printable molds.
%As a side note, the lower shell has two orifices that connect the varying position of the papillary muscles to the bottom side of the shell. These orifices are stenciled in the image before running marching cubes. 

%\subsection{Generation of Individualized Annuloplasty Rings}
%\label{label:annuloplasty}
%Registration to t=0
%Averaging
%Normalizing Position
%Tubing
%Stitching Holes along ring center

\subsection{3D-Printing and Silicone Injection}

The lower and upper shell and the valve holders are sliced in the freely available Ultimaker CURA software and 3D-printed. For all parts except of the lower shell a common rigid filament can be used, e.g.\ polylactic acid (PLA) or ABS. However, one of the key conceptual points is that we use water-soluble polyvinyl alcohol (PVA) for printing of the lower shell. Recent low-cost 3D printer with dual extrusion, such as the Ultimaker3 (Ultimaker B.V., Geldermalsen, The Netherlands) employ this material for support structures (R6). We use it as main material for printing of the lower shell, which is dissolved in water after silicone injection.
%The structure of interest is printed in another material, such as PLA. After printing, the model can be dissolved in (warm) water to wash away the support material. 
We decided to use silicone early on, as other groups have made good experiences with it \cite{Ginty2018,Scanlan2017} and it is relatively easy to handle. Different silicone mixtures exist (e.g.\ with different shore indices) and the suitable mixture for mitral valve repair was assessed experimentally. 

As a side note, in the previous work \cite{EngelhardtBVM2018}, silicone was cast. This is suitable for creation of leaflet models. However, to fill the thin tubes of the chordae tendinae, more pressure is  employable with a syringe. Several specific issues are to mention about the developed injection process. If not done in the way described in the following, large air bubbles remain in the mold or the mold is not filled completely with material. The silicone material has to be stirred and drawn up in a syringe with luer lock. The syringe is closed airtight and a vacuum is created by further drawing up the syringe's piston. This procedure is conducted for approx. 30s to completely degas the silicone. 
Subsequently, the connected mold parts are fixated with a C-clamp, placed upside down and half of the silicone is injected through the orifice at the papillary muscles. This will fill up the subvalvular apparatus. Then, the lower orifices are sealed; the mold is turned around and placed in a 45 $^\circ$ angle. The orifice opposite to the reservoir (Fig. \ref{fig:CAD}) is chosen for injection and as much silicone is inserted until silicone oozes out the orifice at the opposite side (Fig. \ref{fig:Spritz}). Finally, the orifice is sealed and the mold is put to rest in the same $45^\circ$ position. %The reservoir should be completely filled with silicone which might seep down into the mold after 5-10 minutes to fill remaining gaps. 

\section{Experiments}
\label{sec:Exp}

%mention valve thickness

Experiments were conducted to evaluate the material, the chordal thickness, the feasibility and accuracy and surgeons assessed the quality of the models.
\paragraph{Silicone mixture:} For this experiment, a special reusable material test mold was developed and printed in PLA. This mold only consists of the leaflet geometry, but is otherwise similar to the mold we presented in Fig. \ref{fig:CAD}. The subvalvular apparatus was assessed in a separate  experiment. 
 In total, seven different material mixtures were tested (Tab. \ref{tab:silicone}): Erosil 10 (Silikonfabrik.de, Ahrensburg, Germany), Dragon Skin\textsuperscript{TM} 10 Fast and Ecoflex\textsuperscript{TM} 00-10, 00-30 and 00-50 (Smooth-On, Inc., Macungie, Pennsylvania, USA). The goal of this experiment was to find appropriate material for injecting into the gap of the mitral valve leaflets such that 1) the whole gap is filled and 2) no air bubbles remain. Furthermore, the chosen material's viscosity must enable injection molding in general. Viscosity, amount of air bubbles and filling degree was accessed on a 5-point Likert scale. Key surgical factors such as leaflet-like tissue feeling when stitching with authentic suture material and when cutting with scissors were assessed by an expert mitral valve surgeon.

\begin{figure}
  \centering
    \includegraphics[width=1.0\textwidth]{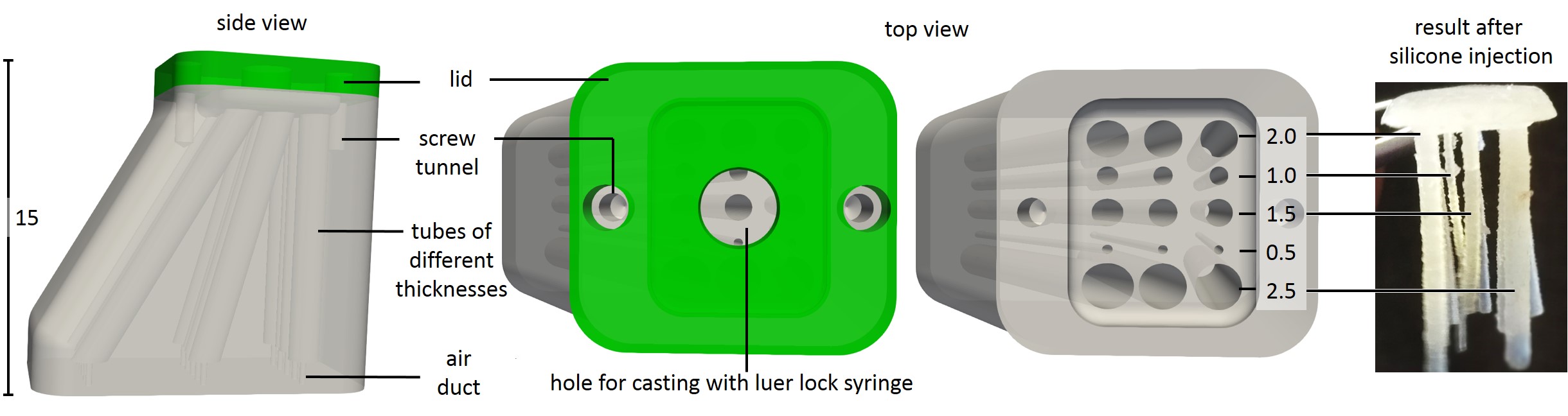}
    \caption{Test object printed in PVA for evaluation of minimal possible chordae thickness.}
\label{Fig:testobject}
\end{figure}

\paragraph{Chordae thickness:}
%Considering the viscosity of the injection material and the accuracy of the printer, it is not clear how thin individual chordae strings can be designed. 
The second experiment was conducted to determine the minimum possible thickness of the individual chordae threads for the chosen material of the previous experiment.
Therefore, a specific test mold object was created with the CAD-program that incorporates a casting hole in a lit (Fig. \ref{Fig:testobject}). This mold furthermore consisted of different tubes with multiple diameters (0.5, 1.0, 1.5, 2.0, 2.5~mm) each having tiny air ducts to release spare gas. The mold was printed from water-soluble PVA. Silicone was injected, the mold dissolved in warm water and it was visually assessed whether the respective tube was filled completely or not.

\paragraph{Feasibility and accuracy in producing different valve geometries from TEE:}

Nine different virtual valve mold geometries were created from TEE acquired with a Philips IE33 X7-2T matrix array transducer (Philips Healthcare, Andover, MA, USA). The data sets contain different pathologies (seven prolapsed valves and two functional mitral valve insufficiency) and degrees of mitral insufficiency (I-III). %The \textcolor{orange}{TEE} volumes were loaded into Philips’ QLab Software and exported in Cartesian DICOM-format.
Valves were segmented and 16 molds were printed using Ultimaker2+ and Ultimaker3 (R6) to produce 16 replica. The production quality, production issues, approximate production time and costs  were assessed. Furthermore, we evaluate the phantom-to-virtual model and the TEE image-to-virtual model accuracy of each valve similarly as in \cite{Scanlan2017}, by measuring different distances: the anterolateral diameter, the posteromedial diameter and the chordae length between leaflet and each papillary muscle tip, using a caliper and virtual measurement tools on a) the phantoms, b) the corresponding virtual models and c) the corresponding TEEs.
%Problems, reproducibility.

\paragraph{Surgical Assessment:}
A group of 12 cardiac surgeons from Heidelberg University Hospital, ranging from surgical residents to mitral valve experts, assessed the realism of the simulator on a 5-point Likert scale %(1 - strongly agree, 2 - agree, 3 - neutral, 4 - disagree, 5 - strongly disagree) 
in a questionnaire. Each of them performed a mitral valve reconstruction on the produced silicone models.
%assessed the suitability and benefit of the simulator for surgical training by performing a mitral valve repair surgery on these valve replica.
The choice of the respective repair techniques was left to the decisions of the surgeons. The group was divided into five expert mitral valve surgeons and seven non-experts. A surgeon was considered to be an expert if his volume of mitral valve reconstructions exceeded 50 open and minimally-invasive operations.
The following categories were assessed: position in simulator, annulus shape, leaflet size, leaflet shape, leaflet thickness, chordae thickness, chordae distribution, chordae length, papillary muscle size, tear resistance at leaflet, cutting leaflet, stitching annulus/leaflet, tear resistance at chordae, tear resistance at papillary muscle, suturing papillary muscle, valve color and valve texture.

\section{Results}

\begin{table}
\resizebox{0.8\textwidth}{!}{%
\caption{Technical criteria for suitability of injection molding for various silicone mixtures were assessed on a 5-point Likert scale (1 - best mark, 5 - worst mork). Surgical properties were evaluated by a mitral valve surgeon.}
\begin{tabular}[t]
{r  c | c c c| c c}
\textbf{Base Material} & \rotatebox{90}{\textbf{ A:B:Slacker}} & \rotatebox{90}{\textbf{Viscosity}} & \rotatebox{90}{\textbf{Air bubbles}} & \rotatebox{90}{\textbf{Filling degree}} & \rotatebox{90}{\textbf{Suturing}} & \rotatebox{90}{\textbf{Cutting}}\\ \hline
  Erosil 10 & 1:1:0 & 2 & 2 & 2 & too solid & ok \\
  Dragon Skin\textsuperscript{TM} 10 Fast & 1:1:0 & 5 & 2 & 4 & good & good\\
  Dragon Skin\textsuperscript{TM} 10 Fast & 1:1:1 & 4 & 2 & 3 & too soft & too soft, sticky\\
  Dragon Skin\textsuperscript{TM} 10 Fast & 1:1:2 & 4 & 3 & 2 & too sticky & too sticky \\
  Ecoflex\textsuperscript{TM} 00-10 & 1:1:0 & 2 & 3 & 2 & too soft & ok\\
  Ecoflex\textsuperscript{TM} 00-30 & 1:1:0 & 1 & 3 & 3 & very good & very good \\
  Ecoflex\textsuperscript{TM} 00-50 & 1:1:0 & 1 & 3 & 3 & very good & very good\\
\end{tabular}}
\label{tab:silicone}
\end{table}

\paragraph{Silicone mixture:} Tab.\ \ref{tab:silicone} summarizes the results of the material tests. Both, Ecoflex\textsuperscript{TM} 00-30 and Ecoflex\textsuperscript{TM} 00-50 (without Slacker) were assessed equally well as leaflet tissue mimicking material. They furthermore got highest ranking in their suitability for injection molding. We use the silicone with a lower Shore index, Ecoflex\textsuperscript{TM} 00-30, for the rest of our experiments. 

\paragraph{Chordae thickness:} 
After dissolving the test object in water, it could be seen that small ducts of less than 1.5~mm were not reliably filled by silicone (Fig.\ \ref{Fig:testobject}). The smallest tubes of diameter 0.5~mm are not represented in the created model at all. This could be due to printing inaccuracies or viscosity of the silicone material. For the rest of the experiments, we set the  chordae thickness to 1.5~mm. This does not fully resemble the fine properties of the chordae strings, however, we want to make sure that the strings are reliably represented in the final valve model. 

\paragraph{Feasibility and accuracy in producing} different valve geometries from TEE: 16 replica from nine different valve models, incorporating the full valvular apparatus of specific patients (R2), were successfully produced. %Dissolving of the lower shell was expedited by breaking away parts of the supporting structures before releasing the material body into water.  
Tab.\ \ref{tab:production} summarizes the production times and costs. They depend on the size of the valve, in particular on the distance of the annulus plane to the papillary muscle tips, as the lower mold's height increases with greater distances. Production took approx. 36h (R7). It could be significantly accelerated by approx. 7 hours by using two 3D printers at the same time and by printing the leaflet holder on stock. The actual producer workload time was around 80 min. 
Material costs were about 16~\euro~for a single valve, depending on the size of the valve. 

Overall production quality was very good and could be maintained over several valves (the upper shell can be re-used, the leaflet holder and lower shell need to be printed for each valve). Visual agreement to the virtual model (R3) is high in all cases. Mean phantom-to-virtual model and the TEE image-to-virtual model distances are $0.87\pm 0.92$~mm and $1.27\pm0.92$~mm for the anterolateral diameter,  $0.44\pm 0.34$~mm and $1.36\pm 0.95$~mm for the posteromedial diameter, $0.70\pm 0.45$~mm and $2.20\pm 2.25$~mm for the distance between leaflet and anterolateral papillary muscle as well as $0.56\pm 0.49$~mm and $1.91\pm 1.22$~mm for the same distance on the posteromedial side.
%\textcolor{red}{Phantom-to-model-to-image distances had a mean standard deviation of 0.85~mm for the anterolateral diameter, 0.76~mm for the posteromedial diameter, 1.34~mm for the distance between leaflet and anterolateral papillary muscle and 1.18~mm for the same distance on the posteromedial side.}
For a few valves, we could identify the following shortcomings: It happened once that the  printed mold had a hole and silicone was erroneously released into a supporting structure compartment. After mold dissolving, this could be fixed by simply cutting away unwanted parts. 
One model had an air bubble at the position of a thin chordae string, leading to a rupture. Minor air bubbles at papillary muscle prolongations and upper portion of the ring fixture appeared, however, they never effected the valve itself. 
%One particular and unusually big virtual valve model was hard to accommodate in the holder. We had to cut parts of the ring fixture to assemble the mold. 

\begin{table}[t]
\resizebox{0.9\textwidth}{!}{%
\centering
\caption{Approximate production times and costs for a single valve. *can be reused, **can be produced on stock independent of the valve geometry }
\begin{tabular}{ | r | c c c |}
\hline
~ & \multicolumn{3}{c|}{\textbf{Production times}} \\
\hline
Segmentation & \multicolumn{3}{c|}{\cellcolor{gray!10}10 min}  \\
\hline
~  & \textbf{Upper Shell*} & \textbf{Lower Shell} & \textbf{Leaflet Holder**} \\
\hline
Negative mold & \cellcolor{gray!10}15 min &\cellcolor{gray!10} 25 min & -  \\
Slicing &\cellcolor{gray!10} 5 min &\cellcolor{gray!10} 5 min &\cellcolor{gray!10} 2 min  \\
3D printing &\cellcolor{gray!10} 320 min &\cellcolor{gray!10} 540 min &\cellcolor{gray!10} 120 min\\
\hline
Silicone injection & \multicolumn{3}{c|}{\cellcolor{gray!10}15 min}\\
Silicone curing & \multicolumn{3}{c|}{\cellcolor{gray!10}240 min}\\
Water quench & \multicolumn{3}{c|}{\cellcolor{gray!10}720-840 min}\\
\hline
total & \multicolumn{3}{c|}{2137 min $\approx$ 36 h} \\
\hline
\hline
%~ & ~ & ~ & ~ \\
~ & \multicolumn{3}{c|}{\textbf{Production costs}} \\
\hline
Printing material & \cellcolor{gray!10}PLA (\SI{50}{\gram}) $\approx$  3,35 \euro & \cellcolor{gray!10} PVA (\SI{70}{\gram}) $\approx$ 9,35 \euro & \cellcolor{gray!10} PLA (\SI{15}{\gram}) $\approx$ 1,00 \euro \\  
Silicone & \multicolumn{3}{c|}{\cellcolor{gray!10} approx. \SI{35}{\gram} $\approx$ 1,70 \euro} \\
\hline
total & \multicolumn{3}{c|}{\cellcolor{gray!10} $\approx$ 16 \euro} \\
\hline
\end{tabular}}
\label{tab:production}
\end{table}

\paragraph{Surgical Assessment:}

\begin{figure}
\floatbox[{\capbeside\thisfloatsetup{capbesideposition={right,top},capbesidewidth=4cm}}]{figure}[\FBwidth]
{\caption{Upper row: Top and side view of two valves before surgical training. Lower row: Surgeons performing annuloplasty, neo-chordae and chordae-loop implantation on the valves fixed in the simulator.}\label{fig:training}}
{\includegraphics[width=7cm]{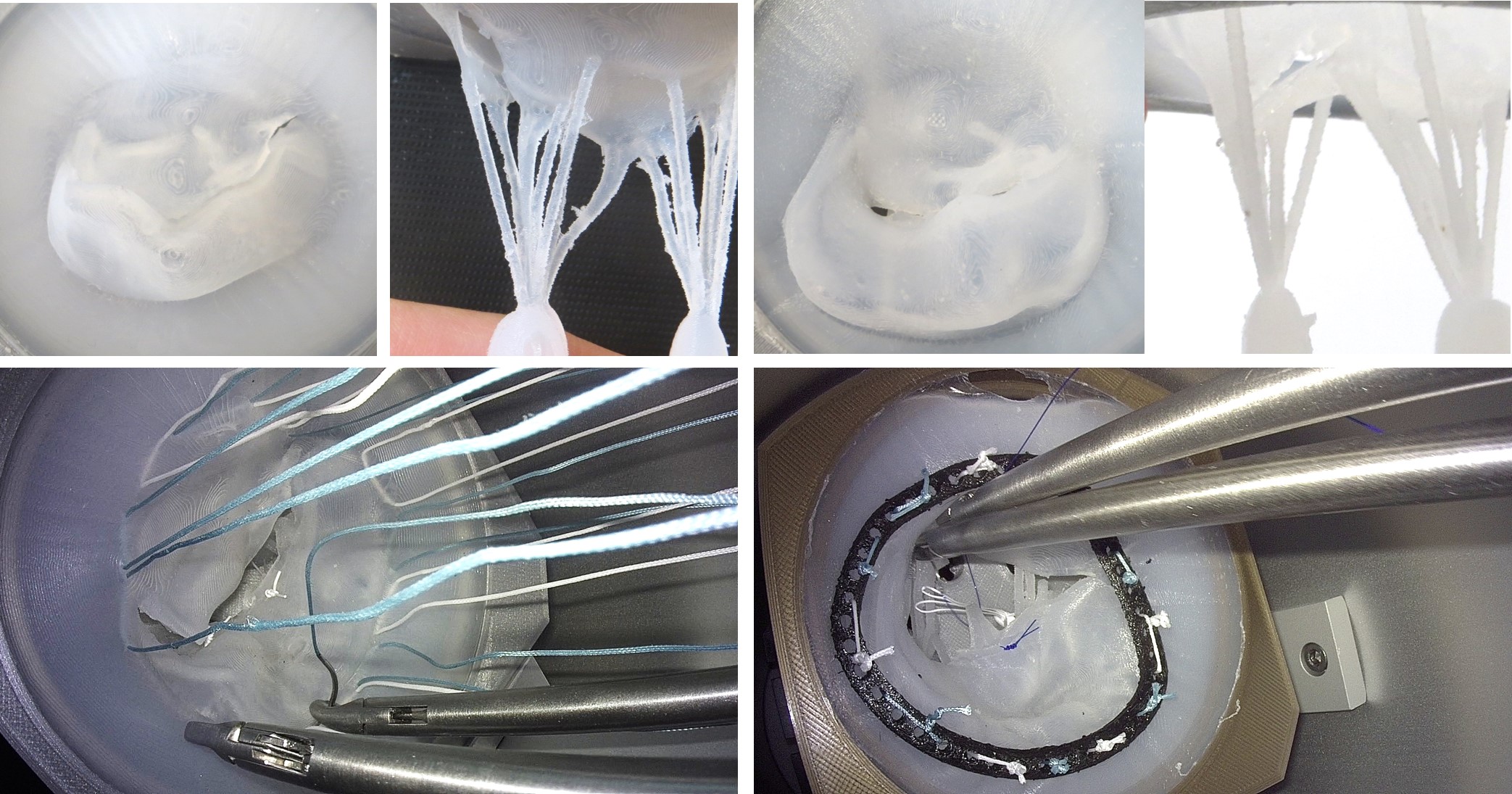}}
\end{figure}

\begin{figure}
  \centering
    \includegraphics[width=\textwidth]{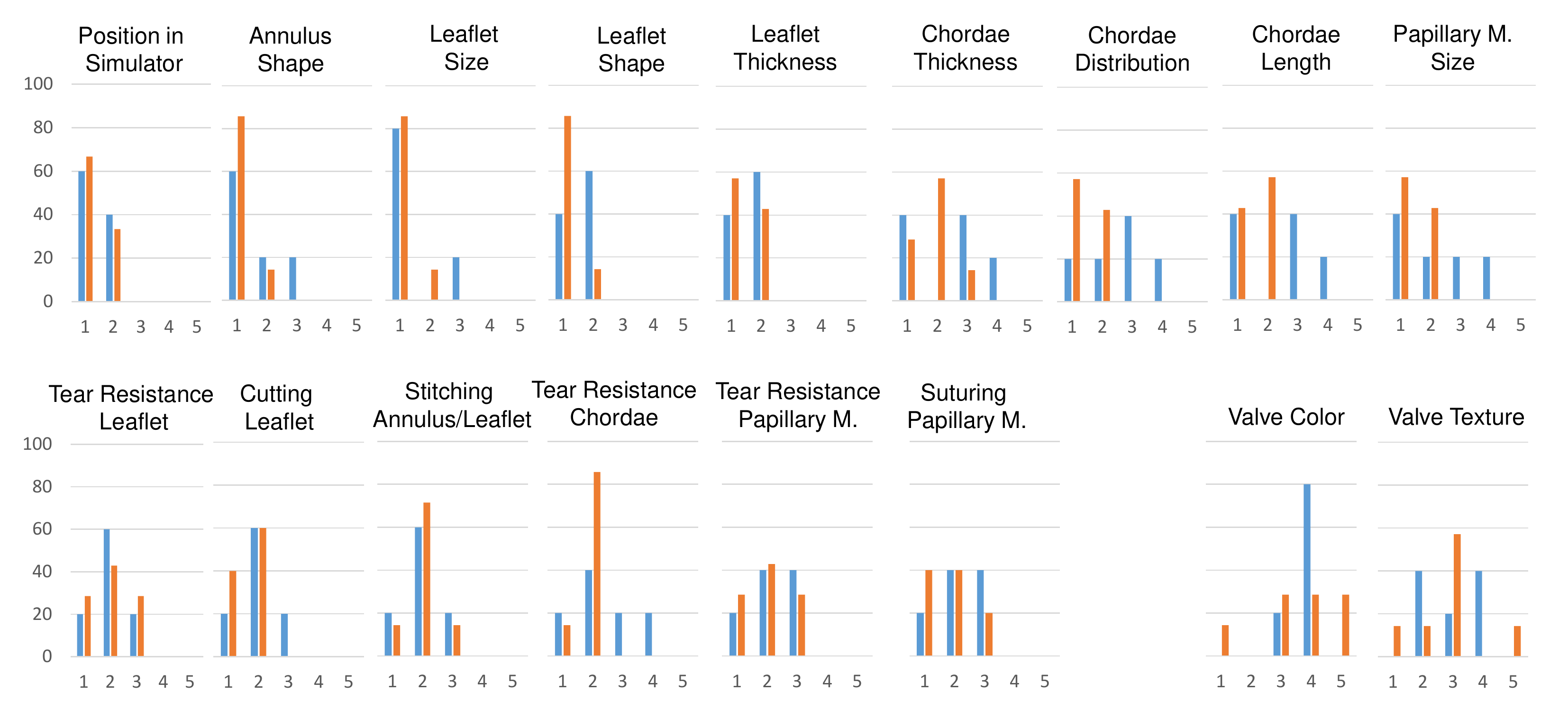}
    \caption{Five expert surgeons (blue) and seven non-experts (orange) assessed the realism on a 5-point Likert scale (1-strongly agree, 2-agree, 3-neutral, 4-disagree, 5-strongly disagree).}
\label{Fig:diagram}
\end{figure}

Fig.\ \ref{fig:training} and the supplemental video show examples of silicone replica and different procedures conducted on these models, e.g. annuloplasty, neo-chordae implantation, chordae-loop implantation.  The results of the questionnaire are presented in Fig.\  \ref{Fig:diagram}. 
 Models maintain a high realism during haptic interaction with instruments and suture material. The realism of leaflet and the annulus (shape, size, thickness) was assessed more positive than the subvalvular apparatus. This was expected, since, due to manufacturing constraints, these structures can not be replicated to reach full satisfaction. The worst vote relates to the realism of the color (median 4) and the texture (median 3) of the material.
 
%\begin{figure}
%  \centering
%    \includegraphics[width=0.6\textwidth]{Training.jpg}
%     \caption{Upper row: Top and side view of two valves before surgical training. Lower row: Surgeons performing annuloplasty, neo-chordae and chordae-loop implantation on the valves fixed in the simulator.}
%     \label{fig:training}
%\end{figure}

\section{Discussion}

The production of individualized high fidelity and flexible models from TEE, which, for the first time, also include the chordae tendineae attached to the papillary muscles, is feasible with our approach. The included chordae maintain a realistic leaflet-to-papillary muscle distance and constrain the moving space of the surgical instruments. The standardized production process guarantees a high anatomical recapitulation of the silicone valves to the segmented models. From a medical point of view, the silicone models are extremely useful for training and patient-individualized rehearsal before surgery, as thoroughly discussed in our related medical publication \cite{EngelhardtEJCTS2018}.
Apart from its benefits for surgical training, the presented method is employable for benchmarking novel technology in development \cite{EngelhardtATS} or potentially, after some adjustments, for investigating catheter-based techniques.
Since the most valid type of evaluation of techniques for valvular interventions is the assessment in realistic hemodynamic environments, we plan the integration of the valves into a flow simulator, exploiting the echogenicity of the silicone material \cite{EngelhardtEJCTS2018}. This will require a comprehensive re-assessment of the currently used soft materials with regard to their suitability for mimicking realistic valve functions. However, it is worth mentioning that Ginty et al.\ \cite{Ginty2018} used the same soft material %(Ecoflex\textsuperscript{\texttrademark} 00-30) 
in their dynamic valve simulator. %Furthermore, the anchoring concept of the valve must be adjusted.} 
\textit{In-vitro} cardiac tissue models are becoming increasingly important \cite{Mathur2015}, but are still out of reach to be employed for surgical training. As a surrogate, porcine hearts are often used (e.g. \cite{Ramphal2005}).
 
%Pseudo-Chordae Distribution:

The current resolution of typical image modalities is not high enough to acquire the thin chordal tree. Researchers lately captured the exercised ovine mitral valve with Micro-CT \cite{Sacks2013} or an exercised porcine mitral valve with 7T MRT \cite{Stephens2017} to gain information on its structure. Having improved imaging data would be a valuable asset for rendering patient-specific valves. A limitation is that the chordae strength is not perfectly resembled in our models owing to the relative soft silicone. In principle, our proposed concept would easily allow usage of two different silicone materials due to the multiple injection holes: A harder silicone type could be injected in the lower shell to fill the subvalvular apparatus first, whereas a softer silicone could be used subsequently for the leaflets.
%Related publications, other anatomical drawings and endoscopic videos served as orientation for development of an algorithm that generates a realistic chordae distribution. 
%According to the surgeons, the manufactured chordal thickness of the phantoms \textcolor{orange}{(assessed as too thick in our evaluation)}, was not considered as a crucial factor for surgical training. Much more relevant were the distances between the papillary muscles and the leaflet attachment for them. 
%WIEDER EINKOMMENTIEREN
Another limiting factor is the current \textit{phantom color} and \textit{texture}. %, which were least convincing (Fig. \ref{Fig:diagram}). %To overcome this drawback, most groups add color pigments to the silicone mixture. This can provide a more realistic tone, but heterogeneous textures, specularities or blood can not be simulated.  
In a related work \cite{EngelhardtMICCAI}, we were able to show how the realism of the silicone model appearance could be significantly increased using a generative adversarial network called tempCycleGAN, which was trained on video sequences of simulated and real mitral valve repairs. The predicted frames replace artificial looking parts of the simulator by more realistic appearances. 
Our vision is that the surgeon will perceive these `hyperrealistic' frames on the endoscopic monitor while operating on the phantom, increasing photo-realism of the surgical target.
%Other material 
%High Impact Polystyrene (HIPS) is a dissolvable support material that is commonly used with ABS. When being used as a support material, HIPS can be dissolved in d-Limonene.
%\paragraph{Funding}
%The research was supported by the German Research Foundation DFG project 398787259, DE 2131/2-1 and EN 1197/2-1.
%\begin{acknowledgements}
%The authors thank Thomas Rosenburg, Heiko Dorwarth from University Magdeburg, Volker Wunsch from University of Applied Science and Hannes Kenngott from University Hospital Heidelberg for their support in 3D-printing. 
%\end{acknowledgements}

\paragraph{Compliance with Ethical Standards}
\paragraph{Conflict of interest} 
The authors declare that they have no conflict of interest.
\paragraph{Ethical standard} 
All procedures performed in studies involving human participants were in accordance with the ethical standards of the institutional and/or national research committee and with the 1964 Helsinki Declaration and its later amendments or comparable ethical standards.
%For this type of study, formal consent is not required.
\paragraph{Informed Consent} 
Informed consent was obtained from all individual participants included in the study.
%For this type of study, formal consent is not required.
\bibliographystyle{spmpsci}      % mathematics and physical sciences
\bibliography{literature}   % name your BibTeX data base


\end{document}